\documentclass[aps,prx,twocolumn,amsmath,amssymb,floatfix,superscriptaddress]{revtex4-2}
\usepackage[latin9,utf8]{inputenc}
\usepackage{stmaryrd}
\usepackage{graphicx}
\usepackage{subfigure}
\usepackage{multirow}
\usepackage{mathtools}
\usepackage{xcolor}
\usepackage{textcomp}
\usepackage{gensymb}
\usepackage{calligra}
\usepackage{braket}
\usepackage{float}
\usepackage{bbold}
\usepackage[hidelinks,colorlinks,
citecolor=darkblue,
linkcolor=darkblue,
urlcolor=nightblue]{hyperref}
\usepackage{amsmath}
\pdfoutput=1

\usepackage{xcolor,mathrsfs,dsfont}
\definecolor{darkblue}{RGB}{0,0,150}
\definecolor{nightblue}{RGB}{0,0,100}
\definecolor{evergreen}{RGB}{0,120,0}
    
\usepackage{graphicx,mathtools,bm}
\graphicspath{{./Figures/}}

\let \oldbm \bm
\renewcommand{\vec}[1]{\oldbm{#1}}

\usepackage[english]{babel}
\usepackage[babel,kerning=true,spacing=true]{microtype}
\usepackage[utf8]{inputenc}

\def\bm{{\vec m}}

\makeatother

\begin{document}
\title{Defect-Bound Excitons in Topological Materials}
\author{Roni Majlin Skiff}
\affiliation{Raymond and Beverly Sackler School of Physics and Astronomy,
Tel-Aviv University, Tel-Aviv 69978, Israel}
\author{Sivan Refaely-Abramson}
\affiliation{Department of Molecular Chemistry and Materials Science, Weizmann Institute of Science, Rehovot 7610001, Israel}
\author{Raquel Queiroz}
\affiliation{Department of Physics, Columbia University, New York 10027, USA}
\affiliation{Center for Computational Quantum Physics, Flatiron Institute, New York 10010, USA}
\author{Roni Ilan}
\affiliation{Raymond and Beverly Sackler School of Physics and Astronomy,
Tel-Aviv University, Tel-Aviv 69978, Israel}

\date{\today}

\begin{abstract}
Excitons, bound states of electrons and holes, are affected by the properties of the underlying band structure of a material. Defects in lattice systems may trap electronic defect states, to which an electron can be excited to form defect-bound excitons.
Here, we examine the effect of band topology on excitons in systems with a single-site defect. We show that in the topological phase, when robust, in-gap, ring-shaped electronic states appear around defects, the excitons' binding energies are lowered as a result of the wide spatial profile of the defect state. In addition, the excitonic wave functions have distinct shapes that change in order with small changes in the model due to the mixed orbital character of the topological bands. 
Our study therefore sheds new light on the dominant mechanisms that govern the behavior of defect-bound excitons in topological materials. 
\end{abstract}

\maketitle

\section{Introduction}
Excitons, formed by Coulomb-bound electrons and holes, are key quasiparticles studied across diverse fields including chemistry, material science, condensed matter physics, and light–matter interactions \cite{koch2006semiconductor,scholes2006excitons,anantharaman2021exciton,wheeler2013exciton,haug2009quantum}. Interest in the field has grown significantly over the past decade, driven by advances in the fabrication and understanding of two-dimensional and layered materials, as well as the discovery that these systems can host stable, long-lived excitons, making them promising candidates for a range of optical and optoelectronic applications \cite{wilson2021excitons, wang2018colloquium, montblanch2023layered, mueller2018exciton, chernikov2014exciton, jin2018ultrafast, jiang2021interlayer,trovatello2020ultrafast,regan2022emerging}. Complementing experimental studies, advanced numerical methods provide ample information about exciton properties, including quantum solutions for their associated wave functions, energies, and stability, stemming from their binding energies and lifetimes\cite{quintela2022theoretical,xie2021theory, sangalli2021excitons,qiu2013optical, qiu2021signatures, qiu2016screening,qiu2015nonanalyticity,wu2015exciton,esteve2025excitons}.

The optical properties of layered and other quantum materials are expected to be modified when the geometric and topological properties of the electronic bands are pronounced. The existing literature highlights the
effects of quantum geometry and topology on excitonic properties such as their spectrum \cite{garate2011excitons,zhou2015berry,srivastava2015signatures,allocca2018fingerprints}, wave functions \cite{paiva2024shift, davenport2024interaction, jankowski2024excitonic, ying2024flat}, dynamics \cite{chaudhary2021anomalous,cao2021quantum,kwan2021exciton,paiva2024shift,xie2024theory}, selection rules \cite{cao2018unifying,xu2020optically,zhang2018optical}, and the topology of the exciton bands \cite{wu2017topological,kwan2021exciton}. In most of these works, properties are treated on a case-by-case basis and rely on numerical simulations. This is owed to the fact that even in pristine systems, analytical models are complicated to solve without resorting to strong approximations.

Defects- lattice imperfections in solid-state systems, particularly in layered semiconductors- are known to significantly alter the optical response of these materials \cite {koperski2017optical, refaely2018defect,aghajanian2023optical}, giving rise to stable, long-lived excitations
that are desirable for single-photon emitters \cite{mitterreiter2021role, azzam2021prospects, he2015single,koperski2015single,srivastava2015optically,chakraborty2015voltage, tonndorf2015single, linhart2019localized}. 
Recent theoretical studies \cite{queiroz2024ring, slager2015impurity} have revealed intriguing features of defects in topological systems. Topological materials are characterized by a "topological obstruction"- the inability to recombine the Bloch wave functions forming a topological band into a set of orthogonal, symmetric, and exponentially localized Wannier functions \cite{marzari1997maximally, brouder2007exponential, soluyanov2011wannier, marzari2012maximally, bradlyn2017topological, po2018fragile}. This obstruction also implies that the spread of their electronic wave functions in real space has a lower bound \cite{marzari1997maximally, brouder2007exponential, soluyanov2011wannier, li2024constraints}.
As was recently shown in Ref.\cite{queiroz2024ring}, in the presence of strong local defect potentials, topological obstructions further give rise to irremovable defect-bound states residing inside topological gaps. These so-called "ring states"- named after the ring-shaped profile of their wave function surrounding the point defect- are constructed from Bloch states belonging to both the valence and conduction bands, and are topologically protected against any adiabatic and symmetric perturbation of the Hamiltonian.
It is therefore expected that excitons formed from electrons excited into a defect state inside a topological gap should bear hallmarks of the ring state's properties. 

In this work, we demonstrate how excitons bound to ring states are affected by their robustness, spatial width, and mixed orbital character, all of which are direct consequences of the topology of the single-particle band structure. We use a simple canonical two-band tight-binding model, which enables a comparison of trivial and topological gaps, and create defect states by adding a local on-site potential on a single lattice site. The Bethe-Salpeter equation is then employed to find the excitons created by electronic excitations from a band to a lattice defect state. Our findings show that while the stability of the ring state ensures the existence of defect-bound excitonic states, their binding energies are lowered, a feature we ascribe to the wide spatial profile of the ring state. Additionally, in the topological case, small changes in model parameters can permute the order of wave functions in the excitonic ladder of states. This behavior is attributed to the mixed orbital character of the bands.

The rest of the paper is organized as follows. In Section (\ref{sec:SPHandES}), we explain the models and the details of the numerical calculations used to find the excitonic properties. Next, we present the results of our calculations: Section (\ref{sec:exciton binding energy}) discusses the exciton binding energies and their dependence on the localization of the defect-state electronic wave function, and Section (\ref{sec:excitonwavefunctions}) explores the different behaviors of excitonic wave functions in trivial and topological systems. Finally, we summarize the results and discuss future prospects in Section (\ref{sec:summary}).

\section{Single-particle Hamiltonian and the Bethe-Salpeter equation}
\label{sec:SPHandES}

As a prototypical example, we consider the Qi-Wu-Zhang (QWZ) model \cite{qi2006topological,asboth2016short} (or half-BHZ model \cite{bernevig2006quantum}) of electrons hopping on a two-dimensional square lattice with two orbitals per site and no spin. The model has two parameters, $t$ and $M$, representing the hopping matrix element and the mass, respectively. The ratio between them controls the transitions between the Chern insulator and trivial phases of the system. The Hamiltonian of the uniform system is 

\begin{equation}
    \begin{aligned}
      \text{\ensuremath{H}}(k_{x},k_{y})&=(\text{\ensuremath{M}}+t(\cos(k_{x})+\cos(k_{y})))\text{\ensuremath{\sigma}}_{\text{\ensuremath{z}}}\\&+t\sin(k_{x})\text{\ensuremath{\sigma}}_{\text{\ensuremath{x}}}+t\sin(k_{y})\text{\ensuremath{\sigma}}_{\text{\ensuremath{y}}}  
    \end{aligned}
    \label{eq.k-sapceH}
\end{equation}
where $\hbar \textbf{k}=\hbar (k_{x},k_{y})$ is the two-dimensional crystal momentum, $\hbar$ is Plank's constant. 
The trivial phase appears for $M<-2t$ and $M>2t$, where the Chern number of both bands of the spectrum is zero, and for $-2t<M<2t$ the phase is topological, namely, the two bands have a non-trivial Chern number equal to plus or minus one. 

Introducing a single-site defect breaks translation symmetry. To do so, we transform the Hamiltonian to real space and insert a perturbation to the lattice, as an on-site potential in one orbital on a single site. The Hamiltonian has the real-space form
\begin{equation}
    \begin{aligned}
    H&=\sum_{m_{x},m_{y}}[M\textbf{c}_{m_{x},m_{y}}^{\dagger}\sigma_{z} \textbf{c}_{m_{x},m_{y}}\\
    &+t\textbf{c}_{m_{x}+1,m_{y}}^{\dagger} \left(\frac{\sigma_{z}+i\sigma_{x}}{2}\right) \textbf{c}_{m_{x},m_{y}}\\
    &+t\textbf{c}_{m_{x},m_{y}+1}^{\dagger} \left(\frac{\sigma_{z}+i\sigma_{y}}{2}\right) \textbf{c}_{m_{x},m_{y}}+c.c]\\
    &+V\textbf{c}_{d_{x},d_{y}}^{\dagger} \left(\frac{\sigma_{0}\pm \sigma_{z}}{2}\right) \textbf{c}_{d_{x},d_{y}}
    \end{aligned}.
    \label{eq.real-spaceH}
\end{equation}
Here, $V$ is the on-site potential of the defect, and 
$\textbf{c}_{m_{x},m_{y}}^{\dagger}=\left(c_{m_{x},m_{y},s}^{\dagger},c_{m_{x},m_{y},p}^{\dagger}\right)$,
where $c_{m_{x},m_{y},\alpha}^{\dagger}$ creates an electron in a lattice site at $\textbf{R}=m_{x}a\hat{x}+m_{y}a\hat{y}$ ($a$ is the lattice constant) in orbital $\alpha$ ($\alpha=s,p$). The vector $\textbf{R}_d=d_{x}a\hat{x}+d_{y}a\hat{y}$ represents the location of the defect, which is chosen to be in the middle of the square lattice, preserving the four-fold symmetry of the system. Hence, the number of lattice points in each spatial direction is odd. The two components of $\textbf{c}_{m_{x},m_{y}}^{\dagger}$ represent two on-site orbital degrees of freedom, which we label $\vert s\rangle,\vert p\rangle$, where $\sigma_{z}\vert s\rangle=+\vert s\rangle,\sigma_{z}\vert p\rangle=-\vert p\rangle$. The last term of the Hamiltonian (\ref{eq.real-spaceH}) therefore represents an on-site defect either on orbital $\vert s\rangle$ or $\vert p\rangle$. 
In all calculations presented, we use periodic boundary conditions to eliminate the effects of edge states.

\begin{figure}
    \centering
    \includegraphics[width=1.0\linewidth]{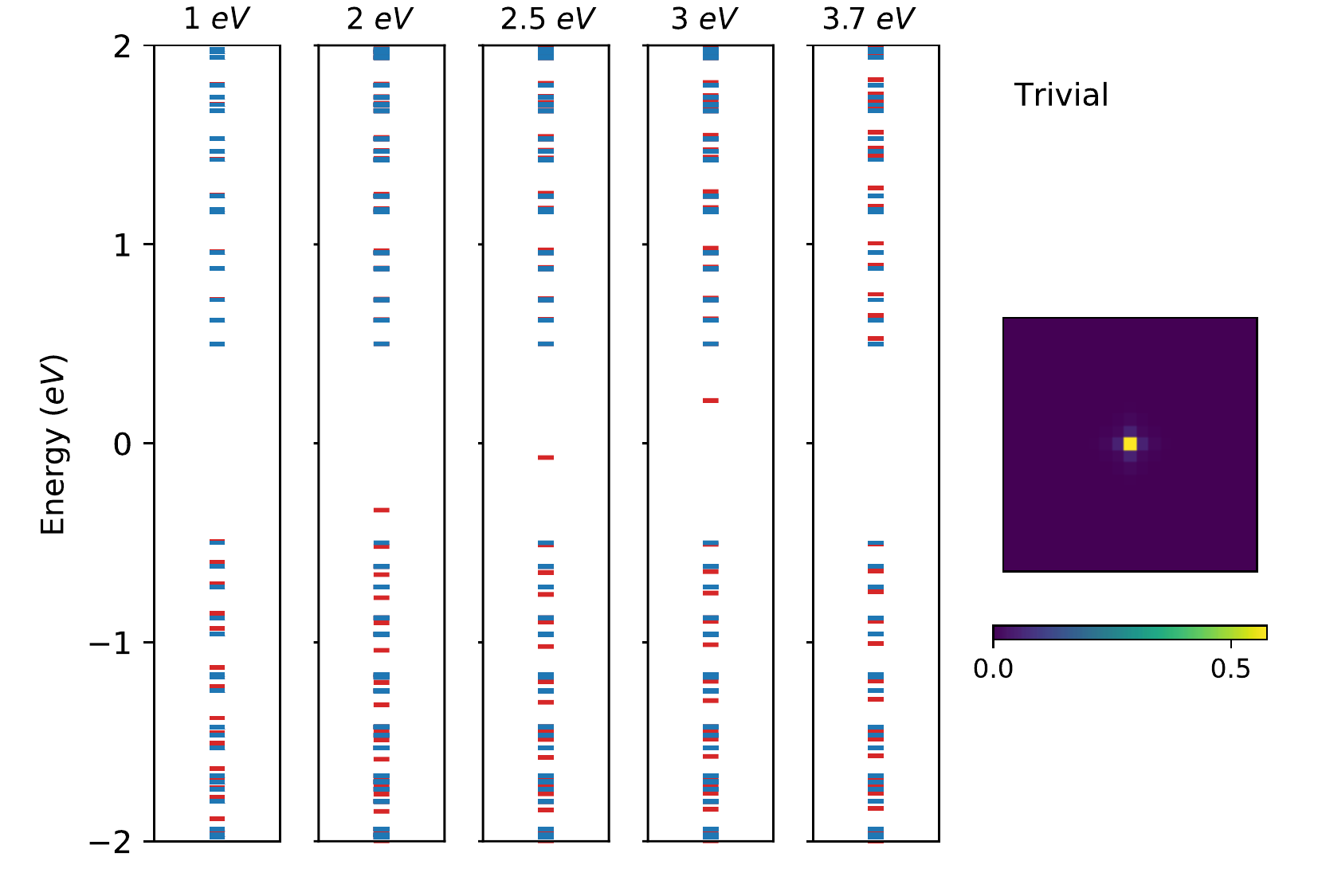}
    \includegraphics[width=1.0\linewidth]{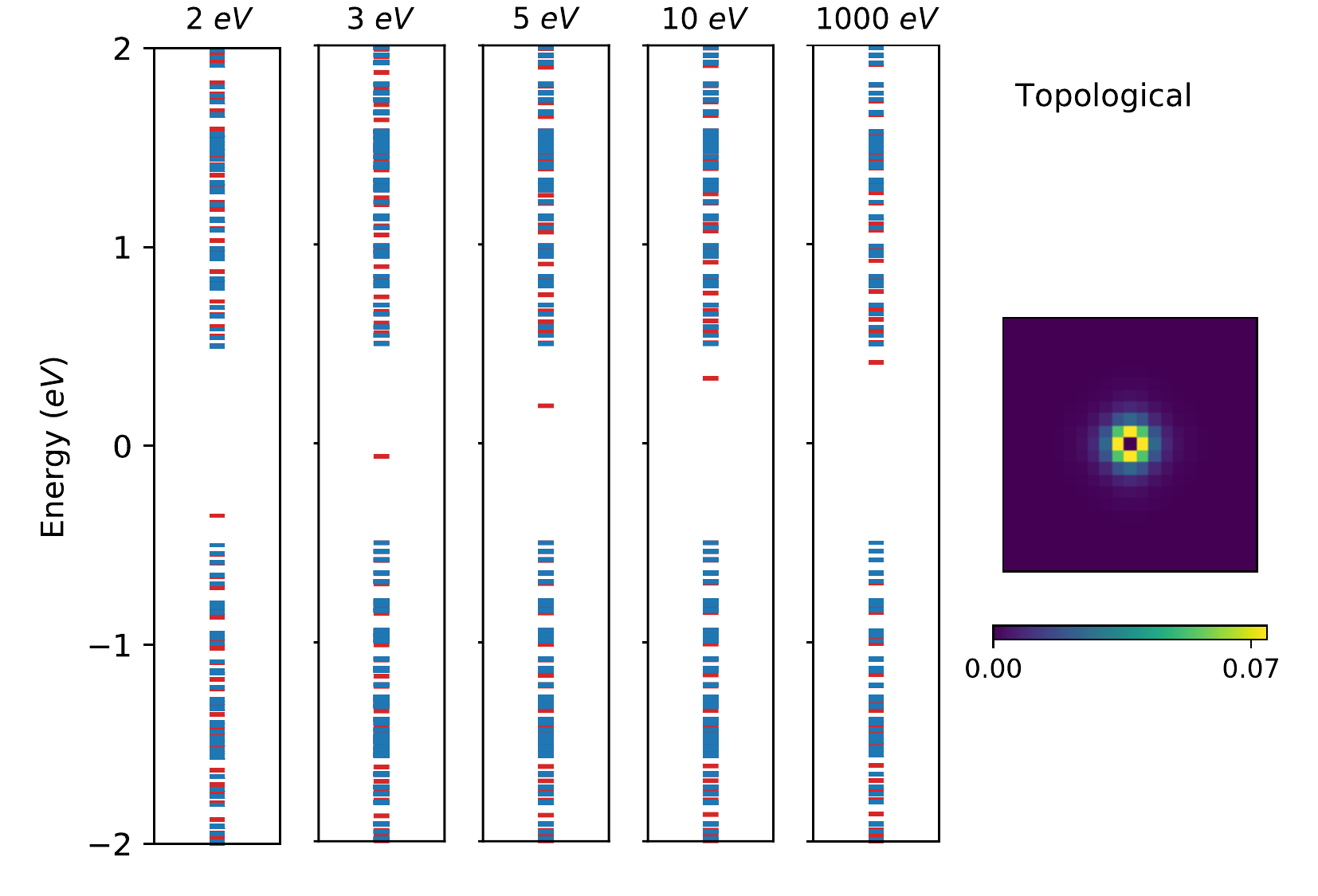}
    \caption{Red lines- Spectra of Hamiltonian (\ref{eq.real-spaceH}) on a lattice of 21x21 sites, with parameters $M=-2.5eV$,$t=1eV$ (\textbf{top left}, trivial systems) and $M=-1.5eV$,$t=1eV$ (\textbf{bottom left}, topological systems) with varying on-site potentials (denoted above each graph) in the $s$ orbital. Blue lines- energy spectra of the same lattices without a defect, for comparison. As can be seen, in the trivial case, an in-gap state appears due to the added potential around $V=2eV$ and merges into the bulk band around $V=3.7eV$. In the topological case, for increasing potentials an in-gap state appears and saturates to a finite energy value of $E=0.408eV$, this is the ring state. Wave function of the in-gap defect state in the trivial case (\textbf{top right}) appearing at energy $E=-0.07eV$ for on-site potential of $V=2.5eV$, with a localized shape on the defect site, and in the topological case (\textbf{bottom right}) at energy  $E=0.408eV$ for large potential values, displaying a ring structure.}
    \label{fig:DefectLocaionInGap}
\end{figure}

We diagonalize the Hamiltonian with the defect (Equation (\ref{eq.real-spaceH})) for different sets of parameters, as exemplified in Figure \ref{fig:DefectLocaionInGap}. Consistent with the general theory of localized perturbations in periodic systems \cite{economou2006green}, the numerical results show that the energy spectrum remains mostly the same compared to the periodic system's spectrum, apart from discrete defect energy levels that may appear. While some bulk wave functions remain unchanged, others are affected by the perturbation and are enhanced around the defect, forming states known as resonances. These resonant states are slightly shifted in energy compared to their energies in the perfect system. In our calculations, the spectrum contains mostly three-fold or higher degenerate energy levels, in addition to non-degenerate energy levels \footnote{In the perfect system with open boundary conditions, all energy levels appear non-degenerate. When the perturbation is added, the resonance eigenstates' energies are slightly shifted compared to their values in the perfect system, while the rest of the energies in the band continuum remain unchanged.}.
 
The appearance and behavior of the discrete defect states depend drastically on the system's topology, as was presented in Ref.\cite{queiroz2024ring} and confirmed by our numerical calculations. In the trivial phase, an in-gap defect state may appear when fine-tuning the defect on-site potential for specific values. This state is unstable and sensitive to small changes in the potential. Figure  \ref{fig:DefectLocaionInGap} (top) demonstrates how the state flows between the bands until it merges with the band continuum. Whether the state flows from the valence to the conduction band or vice versa depends on the orbital affected and on the sign of the on-site potential (attractive/repulsive). The wave function of this in-gap state is point-like and localized on the defect site.
In the topological phase, adding an on-site potential will result in the appearance of an in-gap defect state- the ring state, as was predicted and described in Ref.\cite{queiroz2024ring}, and demonstrated in Figure \ref{fig:DefectLocaionInGap} (bottom). In sharp contrast to the trivial case, this state is stable and remains inside the gap for increasingly large potentials, saturating to a finite energy value. Its wave function has zero amplitude on the defect site and a ring shape around it \footnote{For strong local potentials, a discrete defect state appears at high energies approaching the potential $E_{defect} \rightarrow V$ outside the band continuum (outside the energy range shown in Figure \ref{fig:DefectLocaionInGap}).  In topological systems it appears along with the existence of the in-gap ring state. In trivial systems, it will be the only defect state appearing. In both cases, its wave function is localized on the defect site.}.

Next, we use the calculated single-particle energies and wave functions to find the defect-bound excitonic states created in the system. 
Exciton wave functions are represented as a superposition of electron-hole pairs. Here, we restrict the excitonic basis to transitions between the valence band and the in-gap defect state, in order to isolate the contribution of the localized defect to the excitonic properties. This approximation is motivated by recent experiments and \textit{ab initio} calculations showing that the low-energy optical features in defected transition-metal dichalcogenide monolayers and heterostructures are dominated by such transitions \cite{refaely2018defect,mitterreiter2021role,hotger2023spin,hernangomez2023reduced}. The excitonic wave functions are therefore written as
\begin{equation}
    \Psi_{S}(r_e,r_h)=\sum_{v}A_{S,v}\psi_{d}(r_e)\psi_{v}^{*}(r_h),
    \label{eq. exciton wave function}
\end{equation}
where $\psi_{d}$ is the in-gap defect state electron wave function, and $\psi_{v}^{*}$ are a set of hole wave functions belonging to the valence band. The set of coefficients $A_{S,v}$ are found by solving the Bethe Salpeter equation (BSE) \cite{rohlfing1998electron,rohlfing2000electron,reining1010}, given by
\begin{equation*}
    H_{ex}A_{S}=E_{S}A_{S}.
\end{equation*}
Here, $A_{S}$ is a vector of $A_{S,v}$'s and $E_{S}$ is the exciton energy. The exciton Hamiltonian $H_{ex}$ contains two parts,
\begin{equation}
    H_{ex}=(E_{def}-E_{v})\delta_{vv'}+\Theta_{v}^{v'}
    \label{eq.BSE_H}.
\end{equation}
The energies $E_{def}, E_v$ represent single-particle energies of the defect and valence band states, respectively. The interaction kernel $\Theta_{v}^{v'}$ describes the Coulomb interaction with both direct and exchange terms\cite{reining1010}
\begin{equation}
   \Theta_{v}^{v'}=2E_{vv'}-D_{v}^{v'}.
\label{eq.interaction} 
\end{equation}
The direct interaction term
\[
D_{v}^{v'}=\int dr_{1}dr_{2}\psi_{v}^{*}(r_{1})\psi_{v'}(r_{1})\psi_{def}(r_{2})\psi_{def}^{*}(r_{2})V_{sc}(r_{1},r_{2}).
\]
includes a screened Coulomb potential originally derived for thin films~\cite{rytova2018screened, keldysh2024coulomb} and later developed for strictly two-dimensional dielectric materials \cite{cudazzo2011dielectric}, which has been widely used to approximate layered screening effects in excitonic systems\cite{wang2018colloquium,quintela2022theoretical, qiu2016screening}:
\[
V_{sc}(r)=\frac{e^2}{4\alpha_{2D}}\left[H_0\left(\frac{r}{r_0}\right)-Y_0\left(\frac{r}{r_0}\right)\right]
\]
where $H_0$ and $Y_0$ are the Struve and second-kind Bessel functions, respectively, $r$ is the in-plane coordinate, $e$ is the electron charge, $\alpha_{2D}$ is defined as the $2D$ polarizability, and $r_0=2\pi\alpha_{2D}$ is an effective screening length of the material. We use the approximated form of this potential\cite{cudazzo2011dielectric}:
\[
V_{sc}(r)=-\frac{1}{r_0}\left[\ln\left(\frac{r}{r+r_0}\right)+(\gamma-\ln2)e^{-\frac{r}{r_0}}\right]
\]
with $\gamma\approx0.5772$ the Euler's constant.
The exchange interaction term includes a bare coulomb
potential
\[
E_{v}^{v'}=\int dr_{1}dr_{2}\psi_{v}^{*}(r_{1})\psi_{def}(r_{1})\psi_{v'}(r_{2})\psi_{def}^{*}(r_{2})V_{bare}(r_{1},r_{2}).
\]
Solving the BSE numerically, we find the exciton energies and wave functions.
To plot the exciton wave function, we fix the electron coordinate and plot the wave function of the hole\footnote{Since the electronic part of the wave function belongs to the defect state and is represented by a single state, fixing the electron coordinates results in an overall multiplicative factor to the sum over all hole states, and therefore different values of the electron coordinates will only change the amplitude of the exciton wave function.}. 
The Binding energy is defined as the difference between the single-particle gap, which in our system is the energy difference between the defect state and the top of the valence band, $E_{g}^{dv}= E_{defect}-E_{v,max}$, and the exciton energy\cite{ugeda2014giant, chernikov2014exciton}
\footnote{We note that Ref.\cite{hernangomez2023reduced} presents an alternative definition to the exciton binding energy, as the difference between the expectation values of the diagonal and the full BSE Hamiltonian. We calculated the binding energies of the systems presented in this section following their definition. The obtained results are qualitatively similar to Figure \ref{fig:Binding E vs gap} and are presented in the Supplementary materials.}
: 
\begin{equation}
    E_{b,S}=E_{g}^{dv}-E_{S},
    \label{eq. Eb}
\end{equation}
In the supplementary materials, we show an example result obtained from the BSE: the spectrum, wave functions, and binding energies for a representative set of parameters. 

While it is tempting to assume that the interactions will be strongest between the defect states and the resonance states (described in Section \ref{sec:SPHandES}) due to their enhanced amplitude in the vicinity of the defect, this is not the case. The interaction terms are of similar magnitudes between the defect and resonance states and between the defect and other extended states. Additionally, all valence-band eigenstates, including states deep in the band, contribute significantly to the excitonic energy spectrum. 
We therefore include all valence-band eigenstates in our calculation of the excitonic spectrum and wave functions, in contrast to common calculations of exciton properties, which typically consider only states near the band gap.

\section{Results}
\label{sec:results}
\subsection {Exciton binding energy and its dependence on defect wave function localization}
\label{sec:exciton binding energy}

\begin{figure} [h]
    \centering
    \includegraphics[width=1.0\linewidth]{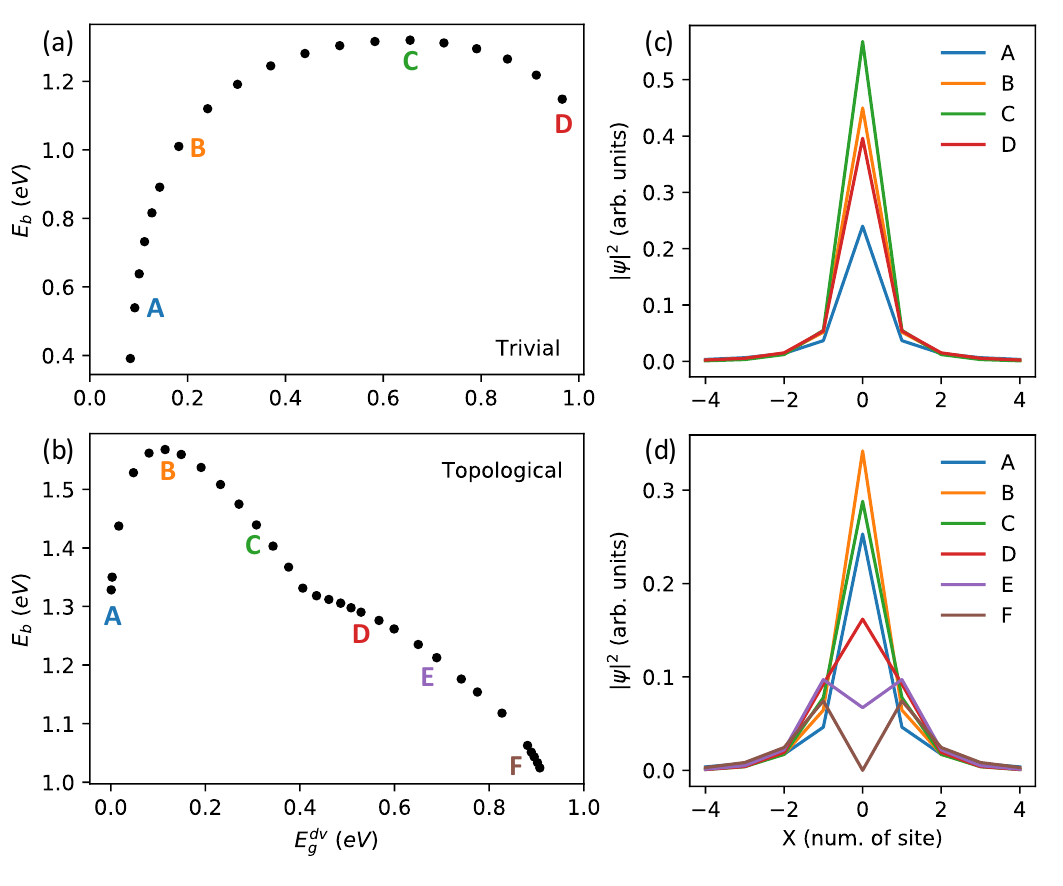}
    \includegraphics[width=1.0\linewidth]{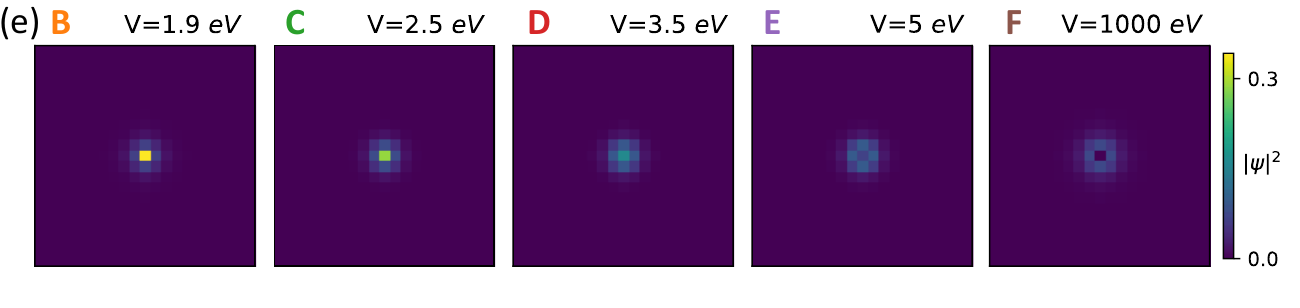}
    \vspace{-0.5cm}
    \caption{Binding energy of the first defect-bound exciton in a trivial system \textbf{(a)} and a topological system \textbf{(b)} with a defect as a function of the gap between the defect state energy and the valence band. Cross-sections of the single-particle defect states' wave functions' amplitude along the $x$-axis in the trivial case \textbf{(c)} and in the topological case \textbf{(d)}, corresponding to various energies at graphs \textbf{(a)} and \textbf{(b)}, respectively (the defect is located at site number 0). As can be seen, the binding energy is lowered as the defect state becomes wider. \textbf{(e)} Two-dimensional plots of the wave functions' amplitudes in \textbf{(d)}. Above each plot is the value of the defect on-site potential that created this state. For large potentials ($V=1000eV$) the defect state is well-saturated into the ring state.
    }
\label{fig:Binding E vs gap}
\end{figure}

As was discussed in the previous section, an in-gap defect state can appear inside the pristine system's band gap, for both trivial and topological systems. In trivial systems it requires fine-tuning of the on-site potential to specific values. In topological systems it is robust and will saturate into a stable energy level for increasingly large potentials\cite{queiroz2024ring}. For both cases, we create a set of systems with varying defect energies by adding and tuning a repulsive potential on the $s$-orbital of one atom. For each system we solve the BSE of an electron excited from the valence band to the defect state, and plot the binding energy of the strongest bound exciton as a function of the energy gap between the defect and the top of the valence band, $E_{g}^{dv}$. The results are presented in Figure \ref{fig:Binding E vs gap}.

In the trivial case \footnote{The parameters used are $M=-2.5$ $eV$, $t=1$ $eV$}, a defect state emerges from the valence band, which has $s$-orbital character. Examining the dependence of the binding energy on the in-gap defect state, we find that it peaks when the defect state's wave function is most localized, see Figure \ref{fig:Binding E vs gap} (a) and (c).

In the topological case \footnote{The parameters used are $M=-1.5$ $eV$, $t=1$ $eV$}, the valence band originates as an $s$-orbital, but a topological band inversion at the $\Gamma$ point ($\textbf{k}=(0,0)$) mixes it with the $p$-orbital conduction band. Turning on the potential, a defect state emerges from the valence band. Examining the dependence of the binding energy on the energy gap, we see that for very small gaps, i.e. when the defect state is slightly above the valence band, a rapid increase in the binding energy is observed. In this regime, the wave function of the defect state is point-shaped and becomes more localized as it parts from the band. This rapid increase is cut off by a sharp drop, which persists monotonically until the defect state saturates to its asymptotic energy value. This behavior is in correlation with the broadening of the defect wave function- first, the point-like shape broadens, then it acquires its even broader ring-shaped profile \cite{queiroz2024ring}, see Figure \ref{fig:Binding E vs gap} \textbf{(d)} and \textbf{(e)}.
Comparing the results of both cases, we see that the binding energy of an exciton bound to a ring state in the topological case ($E_b\approx1.0 eV$) is lower than the binding energy of an exciton bound to deep in-gap states in the trivial case ($E_b\approx1.2 eV$).

\begin{figure}
    \centering
    \includegraphics[width=1.0\linewidth]{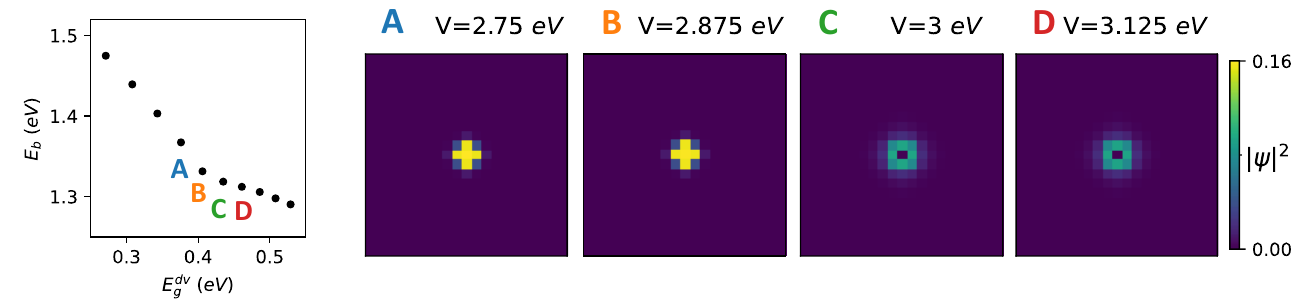}
    \caption{\textbf{Left:} Zoom-in on the kink feature in Figure \ref{fig:Binding E vs gap} \textbf{(b)}. \textbf{Right:} Excitonic wave functions' amplitudes of the strongest bound exciton in each energy marked A-D in the left graph. Above each plot is the value of the defect on-site potential that created this state. An abrupt change in the excitonic wave function's shape is seen where the binding energy graph shows a kink. }
\label{fig:Binding E kink}
\end{figure}

We also note a kink in the graph of the topological case (Figure \ref{fig:Binding E vs gap} \textbf{(b)}), around $E_{g}^{dv} = 0.4eV$. Examining the origin of the kink, no distinct feature in the detect state wave function can be observed at this energy, as the ring state develops a node at higher energies. However, this kink is correlated with a sharp appearance of a node in the excitonic wave function, as can be seen in Figure \ref{fig:Binding E kink}.

According to the above observations, we can associate the behavior of the binding energy with the spread of the wave function of the single-particle state to which the electron is excited to form the exciton. 
In both trivial and topological phases, drifting away from the band continuum localizes the defect state, resulting in more tightly bound excitons.
However, the characteristic behaviors part as the defect state starts to broaden into a ring state due to the topology of the band structure, which results in a lowering of the binding energy of the excitons in the topological phase.  
As the single-particle wave function broadens, the exciton wave function acquires a wider profile as well, however, the behavior of the exciton wave function is more complex, as will be discussed in the next section. 
We note that the bandwidth also influences the exciton binding energy, in addition to the wavefunction spread. However, as shown in the Supplementary Material, the dependence of the binding energy on the bandwidth is very similar in the trivial and topological phases. We therefore conclude that the qualitative difference between the two phases is primarily associated with the increased spatial extent of the topological ring-state wave function.

The connection between the exciton binding energy and its spatial spread was examined before in various systems. Ref.\cite{dvorak2013origin}, for example, analyzed the connection between the localization of three-dimensional excitons and their binding energies in different materials, and found that compact excitons have higher binding energies. In a Graphene-transition-metal dichalcogenides heterostructure, it has been predicted that chalcogen vacancies will give rise to defect-localized excitons with high binding energy compared to these in the pristine system \cite{hernangomez2023reduced}.
Recent studies that discuss excitons in organic semiconductors \cite{jankowski2024excitonic} and in flat band systems \cite{ying2024flat} also show a connection between the spread of the excitonic wave function \cite{haber2023maximally} and the quantum metric, similar to bounds in the single-particle case \cite{marzari1997maximally,marzari2012maximally,brouder2007exponential, soluyanov2011wannier}. Here, we show that in the case of defect-bound excitons, the binding energy is correlated to the width of the defect state wave function, which is directly influenced by the topological obstruction of the bands. \cite{queiroz2024ring} \footnote{An analysis of the binding energy as a function of the ratio between the gap and the bandwidth further supports our findings and can be found in the Supplementary Material.}

\subsection {Exciton wave functions}
\label{sec:excitonwavefunctions}
\begin{figure}
    \centering
    \includegraphics[width=0.8\linewidth]{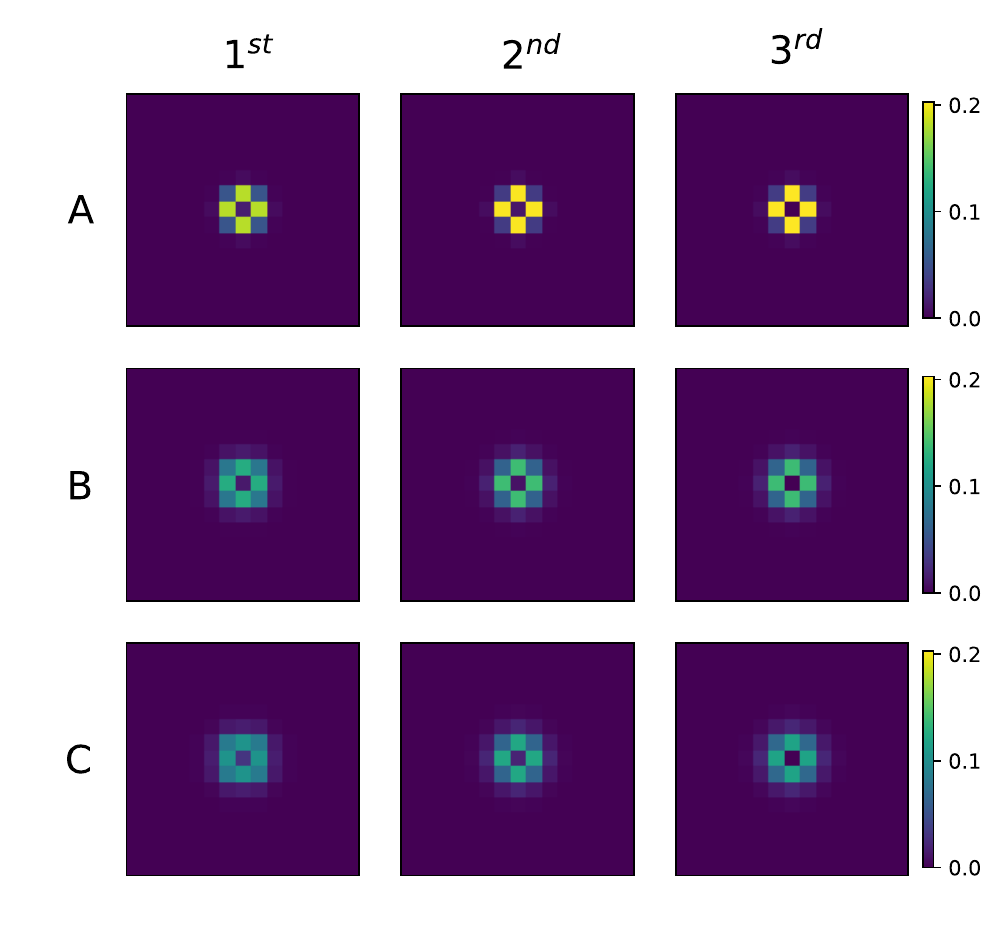}\vspace{-0.5cm}
    \caption{Wave functions of the 1st (strongest bound), 2nd, and 3rd excitons of trivial systems A, B, and C, with a defect in the $s$ orbital. The systems' parameters appear in Table \ref{table:trivial}. As can be seen, in contrast to the topological case, all wave functions are similar in character, and appear in the same order in the excitonic ladder of states.}
\label{fig:trivial wave functions}
\end{figure}

\begin{figure}
    \centering
    \includegraphics[width=0.8\linewidth]{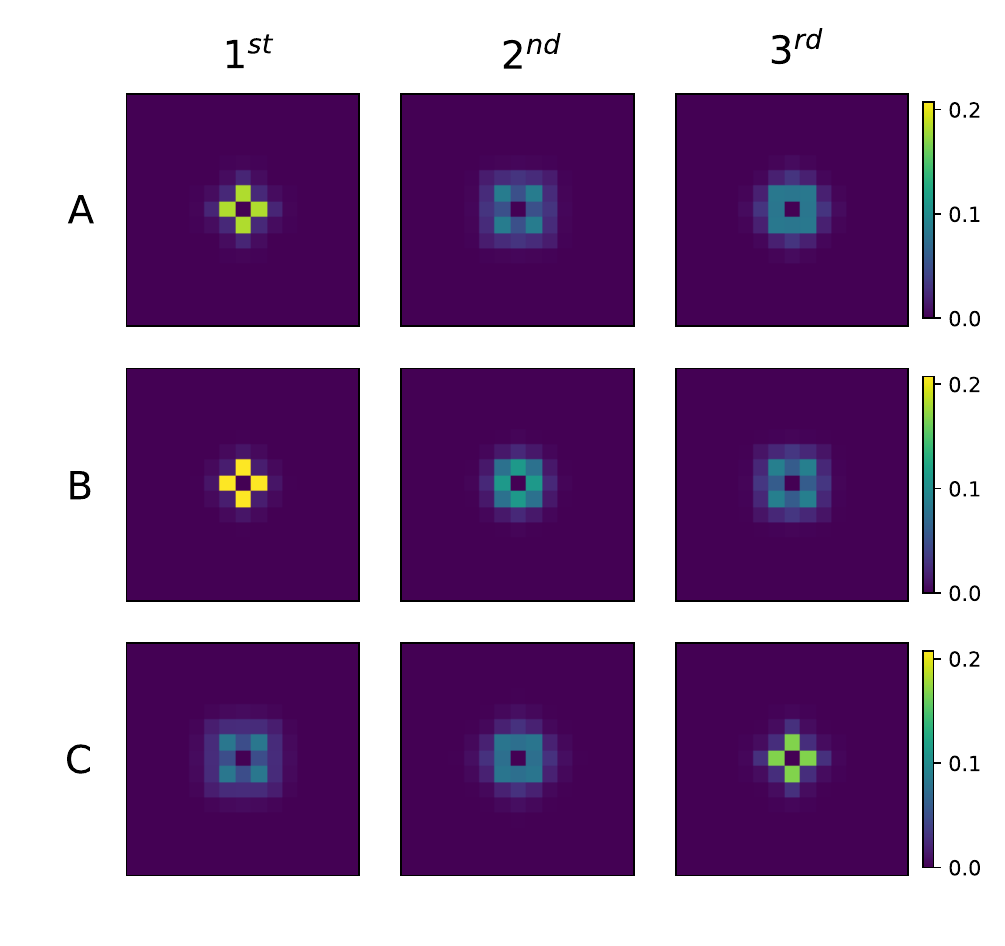} \vspace{-0.5cm}
    \caption{Wave functions of the 1st (strongest bound), 2nd, and 3rd excitons of topological systems A, B, and C, with a defect in the $s$ orbital. The systems' parameters appear in Table \ref{table:topological}. Comparing systems A and C, where the gap between the bands is fixed and the bandwidth changes, the most tightly bound exciton in system A has a form similar to the one that appears as the third exciton in system C. When systems B and C are compared, the ratio between the bandwidth and the gap ($E_{BW}/E_{g}^{dv}$) is fixed but each one of them changes separately, the first and third excitons switch places in the excitonic ladder of states.}
\label{fig:topological wave functions}
\end{figure}

In this section, we examine the excitonic wave functions and their dependence on the model's parameters.
First, we consider the three strongest bound excitons in trivial systems with a defect, as presented in Figure \ref{fig:trivial wave functions} and Table \ref{table:trivial}. We tune the parameters of the Hamiltonian (\ref{eq.real-spaceH}), $M$ and $t$, to control the single-particle band gap $E^{dv}_{g}$, bandwidth $E_{BW}$, and their ratio $E_{BW}/E_{g}^{dv}$. As the model parameters change, we observe that the excitonic wave functions remain qualitatively similar in shape in all systems, and, specifically, that they maintain their order. 
We note that these excitons feature a node in their center, despite the defect state being a localized point-like state without a node. 
Next, we consider topological systems with a defect, see Figure \ref{fig:topological wave functions} and Table \ref{table:topological}. The defect potential is fixed to be strong and repulsive, so that the ring state is saturated and insensitive to small changes in the parameters. Examining the first three excitons reveals that they have qualitatively distinct spatial forms. Furthermore, small changes in the parameters affect the ordering of the wave functions, namely, the first, second, and third exciton states are swapped, see caption of Figure \ref{fig:topological wave functions} for details.

We attribute the complex behavior of the excitonic wave functions in the topological case to the mixed orbital character of the bands. Recall that the exciton wave functions Equation (\ref{eq. exciton wave function}) are constructed from the single-particle wave functions of both valence and defect states. In the trivial case, the conduction band does not play a role in the process.
In contrast, in the topological case, a band inversion causes a mixing of the bands, and the single-particle defect state itself must have a non-trivial projection on both bands\cite{queiroz2024ring}. Therefore, although excitons are formed by Coulomb interaction between the valence band and the in-gap defect state, their formation involves states from the conduction band. The superposition of single-particle states in the exciton wave function (Equation (\ref{eq. exciton wave function})) then reflects this mixed orbital character. Changing the model parameters changes the projection of the defect state onto the two bands, since changing the bandwidth $E_{BW}$ or gap $E^{dv}_{g}$ changes the mixing between the two orbitals in the bands. Changing both of these quantities together but keeping their ratio ($E_{BW}/E_{g}^{dv}$) fixed does not change the relative projections of the ring state onto the two bands, however, the single-particle energies change with respect to the energy scale of the Coulomb interactions. This will modify the weights of the single-particle states in the BSE solutions and, in turn, change the shapes of the excitonic wave functions and their energies. We note that, in the topological regime, both bands possess a mixed orbital character due to the band inversion. Consequently, bulk excitons also inherit this mixed character. However, bulk excitons are superpositions of many electron-hole pairs with weights determined by the BSE coefficients, so the extent to which this mixed orbital character is manifested depends on the particular excitonic state. In contrast, the same mixed-orbital single-particle ring state enters every ring-state-bound exciton. We therefore expect the signatures of the underlying topology to be more robust and pronounced in excitons bound to the ring state.

\begin{table}[h!]
\centering
\begin{tabular}{|c|c|c|c|c|c|c|c|c|c|c|c|} 
\hline
  & $M$ & $t$ & $|M/t|$ & V & $E_{BW}$ & $E_{g}^{cv}$ & $E_{g}^{dv}$ & $\frac{E_{g}^{dv}}{E_{BW}}$ & $E_{B,1}$ & $E_{B,2}$ & $E_{B,3}$ \\ [1ex]
\hline\hline
A & -1.8 & 0.4 & 4.5 & 2.88 & 1.58 & 2 & 1.9 & 1.2 & 2.25 & 2.18 & 2.15 \\ [1ex]
\hline
B & -3.6 & 0.8 & 4.5 & 5.76 & 3.17 & 4 & 3.8 & 1.2 & 1.89 & 1.71 & 1.67 \\ [1ex]
\hline
C & -2.6 & 0.8 & 3.25 & 3.9 & 3.16 & 2 & 1.9 & 1.6 & 1.67 & 1.53 & 1.45 \\ [1ex]
\hline
\end{tabular}
\caption{Parameters of trivial systems A, B, and C appearing in Figure \ref{fig:trivial wave functions}. $M$ and $t$ are the tight-binding parameters appearing in Equation (\ref{eq.real-spaceH}). $V$ is the the on-site potential in each system, tuned such that the defect state is located inside the gap in
approximately the same energy relative to the band gap and bandwidth. $E_{BW}$ is the bandwidth of the valence band, $E_{g}^{cv}$ is the gap between the valence and the conduction bands in the perfect system, $E_{g}^{dv}$ is the gap between the in-gap defect state and the top of the valence band.
The binding energies of each system's first three bound excitons appear in the last three lines. All magnitudes (except ratios) are in $eV$ units.}
\label{table:trivial}
\end{table}

\begin{table}[h!]
\centering
\begin{tabular}{|c|c|c|c|c|c|c|c|c|c|c|} 
\hline
  & $M$ & $t$ & $|M/t|$ & $E_{BW}$ & $E_{g}^{cv}$ & $E_{g}^{dv}$ & $\frac{E_{g}^{dv}}{E_{BW}}$ & $E_{B,1}$ & $E_{B,2}$ & $E_{B,3}$ \\ [1ex]
\hline\hline
A & -1.4 & 1.2 & 1.17 & 2.76 & 2 & 1.55 & 0.56 & 1.64 & 1.62 & 1.60 \\ [1ex]
\hline
B & -0.8 & 0.65 & 1.23 & 1.58 & 1 & 0.8 & 0.5 & 1.78 & 1.71 & 1.68 \\ [1ex]
\hline
C & -1.6 & 1.3 & 1.23 & 3.16 & 2 & 1.59 & 0.5 & 1.49 & 1.45 & 1.44 \\ [1ex]
\hline
\end{tabular}
\caption{Parameters of topological systems A,B and C appearing in Figure \ref{fig:topological wave functions}. Parameter descriptions are the same as in Table \ref{table:trivial}. All magnitudes (except ratios) are in $eV$ units.}
\label{table:topological}
\end{table}

Lastly, we point out that the behavior of the binding energy presented in Figure \ref{fig:Binding E kink} can also be interpreted in the context of the discussion of the mixed orbital character of the bands. There, increasing the on-site potential from $V=2.875$ $eV$ to $V=3$ $eV$ results in an abrupt change in the shape of the strongest bound exciton. This change of the on-site potential causes a slight change in the ring state's energy, but also modifies its projections onto the two bands, which affects the exciton wave function. Such an effect is not observed for a system with a trivial gap.

\section{Summary}
\label{sec:summary}
The connection between excitons and band topology has been extensively explored in recent years. In this work, we add another ingredient- a lattice defect, and compare excitons created in electronic defect states in trivial and topological systems. Our numerical analysis of defect-bound excitons shows that in both cases, the exciton binding energy depends on the spread of the defect state's wave function, however, due to the wide spatial profile of the ring state, the binding energies in topological materials are typically lower. In addition, we show that in topological systems the wave functions of the low-lying excitonic states have distinct shapes, and their order in the ladder of excitonic energies is swapped with small changes in the model's parameters. This is in contrast to the trivial case, in which the wave functions of the low-lying excitonic states are qualitatively similar and ordered across a broad range of parameters. We attribute this behavior of the topological systems to the mixed orbital character of the topological bands. Despite neglecting direct transitions between valence and conduction bands, the conduction band still plays a key role in the exciton formation since the ring state is inevitably constructed from states from both bands. The $k$-space topology is thus manifested in the real-space properties of excitons in a pronounced and complex manner.

In this work, we model lattice defects as single-site, single-orbital on-site potentials, which is an idealized description of impurities in real materials. As discussed in Ref.~\cite{queiroz2024ring}, which introduced the concept of ring states around single-site lattice defects, several extensions of this minimal model can be considered. One natural extension is to include defect potentials acting on multiple orbitals at the same lattice site, as in the case of a lattice vacancy. Our numerical calculations indicate that each orbital affected by the defect potential gives rise to a corresponding ring state. For the parameter range considered here, these ring states generally appear at different energies and therefore exhibit negligible hybridization. At low impurity densities, weak hybridization among defect states leads to multiple in-gap levels near the single-impurity energy, while for high impurity densities or clustered single-site impurities, stronger hybridization causes these levels to approach the spectrum of boundary modes. Although such extensions may modify quantitative details of the excitonic properties, they originate from the same underlying topological character of the bands. We therefore expect the main qualitative conclusions of this work to remain robust.

While single-particle electronic states are measurable using common techniques such as scanning tunneling microscopy (STM), studies show that excitons' wave functions are also accessible in experiments \cite{man2021experimental,dong2021direct,fukutani2021detecting}, even those of defect-bound excitons \cite{matsuda2003near}. We therefore hope that our findings will motivate future experimental work in the field of excitons in quantum and layered materials. From a theoretical perspective, we show that excitons' properties depend on the topology of the entire band, and therefore cannot be captured by simplified models such as effective mass approximations. Our study within the tight-binding framework also highlights the role of atomic orbitals while still providing a relatively simple approach from which the effects of band inversions can be singled out. Our results are expected to shed light on the role of electronic bands' topology in exciton properties, and should therefore be of interest both for future theoretical as well as experimental studies of novel materials. 

\section{Acknowledgments}
The authors thank Tobias Holder, Nicolas Regnault, and Noa Feldman for enlightening discussions.
R.I. and R.M.S. are supported by the U.S.-Israel Binational Science Foundation (BSF) Grant No.2018226 and the Israel Science Foundation (ISF) Grant No. 1790/18 and 2307/24. S.R.A. is supported by a European Research Council (ERC) Grant agreement No. 101041159. R.Q. is supported by the NSF CAREER Grant No. DMR-2340394.  

\bibliographystyle{unsrt}
\bibliography{TopoExcitonBib}

\newpage

\section*{Supplementary material}

\subsection*{Alternative definition of the binding energy}

\begin{figure}[H]
    
    \includegraphics[width=0.8\linewidth]{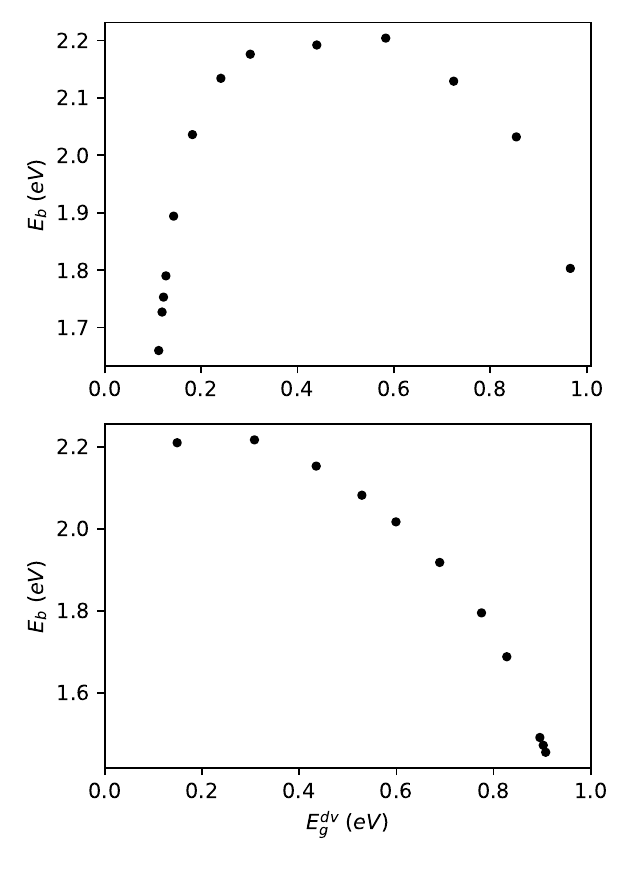}
    \caption{Binding energy, calculated using Equation \ref{eq. Eb extended}, of the strongest bound exciton in a trivial system (upper) and a topological system (lower) with a defect as a function of the gap between the defect state energy and the valence band.}
\label{fig:Binding E Extended}
\end{figure}

In the main text, we use the definition of the exciton binding energy as the difference between the band gap and the exciton energy (Equation (\ref{eq. Eb})) (the lowest exciton energy is also termed the optical gap \cite{ugeda2014giant, dvorak2013origin}). This magnitude represents the energy required to create an exciton. It is easily accessible in experiments, and is commonly used in both experimental and theoretical works. 
Ref.\cite{hernangomez2023reduced} uses an alternative definition for the exciton binding energy, as the difference between the expectation values of the diagonal part of the BSE Hamiltonian (Equation (\ref{eq.BSE_H})) and its full form: 
\begin{equation}
    \begin{aligned}
    E_{b,S}&=\langle  \Psi_{S} \vert H_{ex}-\Theta_{v}^{v'}\vert \Psi_{S} \rangle - \langle \Psi_{S} \vert H_{ex}\vert \Psi_{S} \rangle\\&=\sum_{k}\vert A_{S,k}\vert ^{2}(E_{def}-E_{v}) - E_{S}
    \end{aligned}
    \label{eq. Eb extended}
\end{equation}
This definition can be interpreted as the energy cost to break an exciton into all the electron-hole pairs that constitute it. 
We calculated the binding energies of the systems presented in the main text in section \ref{sec:exciton binding energy} following their definition. The obtained results are qualitatively similar to Figure \ref{fig:Binding E vs gap} and are presented in Figure \ref{fig:Binding E Extended}

\subsection*{Example of Exciton spectrum}

\begin{figure}[H]
    \centering
    \includegraphics[width=0.5\linewidth]{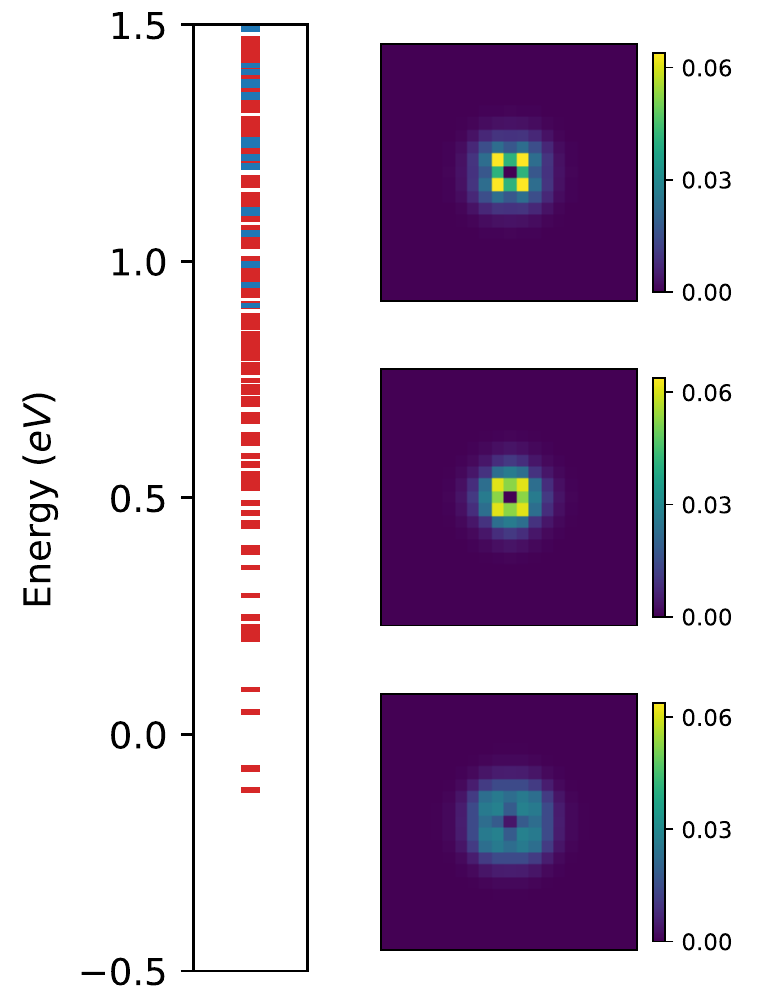}
    \caption{\textbf{Left}: Red lines- Exciton energy spectrum of a system with $M=-1.5$ $eV$ ,$t=1$ $eV$ and a local potential $V=1000$ $eV$ in the $s$ orbital. Blue lines- the energy difference between the defect energy and all single-particle valence energies, See text for description. \textbf{Right}: from top to bottom: wave functions of the first, second and third lowest-energy, strongest bound excitons, with binding energies are $E=1.025, 0.981, 0.977$ $eV$, respectively. The wave functions drawn are of the hole, as the location of the electron does not change their shape.}
\label{fig:BSE solution M=-1.5 spec and wf}
\end{figure}

In Figure \ref{fig:BSE solution M=-1.5 spec and wf} we show an example for exciton solutions for a representative system. In the spectrum, the blue lines represent all single-particle energy gaps between the defect and the valence states: $E_{def}-E_v$. Hence, these were the excitation energies if no interactions were present. Red lines represents the excitonic spectrum, namely the bound two-particle states' energies, created when including the Coulomb interaction, which creates bound states below the single-particle band continuum. We note that in this figure, the lowest exciton energies are negative. This is not a consistent or defining feature of this system, and can be changed by modifying the strength of the potential, for example. 
In addition, we note that in our calculations the numerical ladder of exciton energies does not follow a two-dimensional hydrogen-like spectrum. This is reasonable for a  small systems size that deviate considerably from the continuum problem where this hydrogenic approximation is valid. Also, the potential we use is a screened Coulomb potential, and we consider a system with a defect in one orbital of particular symmetry.

\subsection*{Dependence of binding energy on bandwidth-gap ratio}

To establish further the role of the topological character of the bands, we examine the dependence of the binding energy on the bandwidth of the valence band. In each regime, we choose the model parameters such that the gap between the valence and conduction band is fixed, and tune them such that only the bandwidth changes. Then, we add a repulsive on-site potential, $V$, in the $s$ orbital, to created an in-gap defect state, and we distinguish between the two cases. 

\begin{figure}[H]
    \centering
    \includegraphics[width=0.8\linewidth]{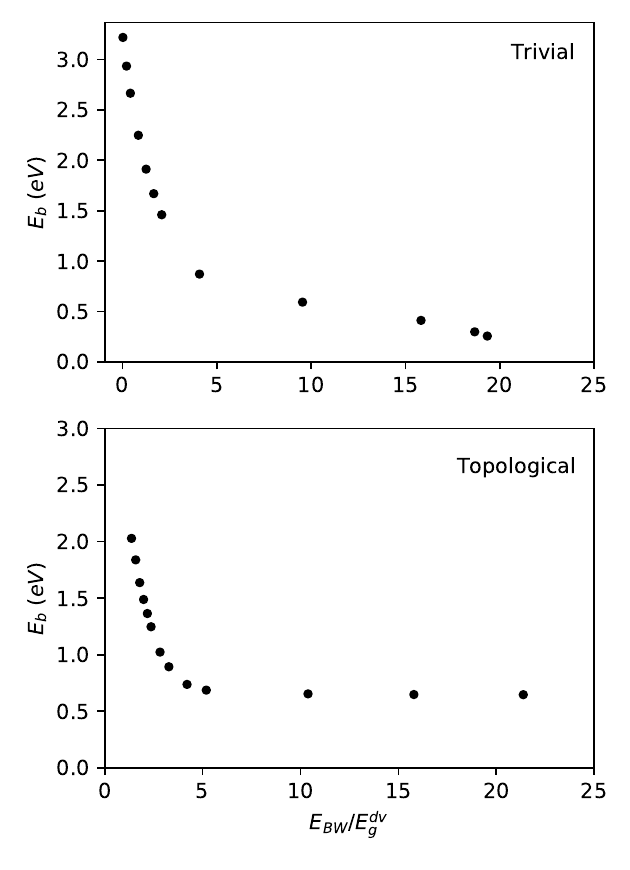}
    \caption{Binding energy of the strongest bound exciton in the spectrum of a trivial system (upper) and topological system (lower) with a defect as a function of the ratio between bandwidth and gap energies.}
    \label{fig:Binding E vs Ebw/Eg}
\end{figure}

In the trivial case, we fine-tune $V$ such that the defect is located in the same energy inside the gap. The value of $V$ changes drastically for every set of parameters (different bandwidth), which emphasizes the sensitivity and instability of the existence of the defect state in that case. The energy of the top of the valence band modifies slightly when the perturbation is inserted, mainly for larger bandwidth, but the overall change in the gap is not significant. In the next step, we consider the binding energy of the lowest bound exciton, $E_b$, as a function of the ratio between the bandwidth, $E_{BW}=E_{v,max}-E_{v,min}$, and the gap from defect state to the top of the valence band, $E_{g}^{dv}$. The results are presented in Figure \ref{fig:Binding E vs Ebw/Eg}. For large values of $E_{BW}/E_{g}^{dv}$, the binding energy is low. It increases as $E_{BW}/E_{g}^{dv}$ is lowered and when this ratio becomes of order unity and below, a rapid increase of the binding energy is observed until the flat band limit is reached.

For the topological system, we fix $V=1000$ $eV$, such that the ring state is well-saturated. We note that the energy of the ring state changes slightly for the different model parameters, namely, its saturation energy depends on the bandwidth of the bands participating in its formation. We then plot again the binding energy as a function of the energy gap between the ring state energy and the valence band. When $E_{BW}/E_{g}^{dv}$ goes below $E_{BW}/E_{g}^{dv}\approx3$, a rapid increase in binding energy is observed. This behavior is similar to the trivial case, as it is clear that the binding energy of the excitons increases as the band flattens, for both trivial and topological bands. However, Bearing in mind that the bandwidth of a topological band has a natural lower bound and cannot be made completely flat, the binding energy cannot keep increasing beyond this point. This is not true for the trivial case for which the atomic limit can be smoothly approached. This is related to the bounds on the spread of the wave functions of the band in the local basis, and is consistent with our previous intuition regarding the effect of the width of the wave function on the binding energy. 

\end{document}